\documentclass[letterpaper,epsf,pra,floatfix,amsmath,twocolumn]{revtex4}

\usepackage{amsmath}
\usepackage{epsfig}
\usepackage{epsf}
\usepackage{array}
\usepackage{verbatim}
\usepackage{hyperref}

\newcommand{\be}{\begin{equation}}
\newcommand{\ee}{\end{equation}}

\newcommand{\rref}[1]{(\ref{#1})}
\newcommand{\eref}[1]{Eq.~(\ref{#1})}
\newcommand{\esref}[1]{Eqs.~(\ref{#1})}

\renewcommand{\Re}{\mathrm{Re}}
\renewcommand{\Im}{\mathrm{Im}}

\newcommand{\ocite}[1]{Ref.~\onlinecite{#1}}

\begin{document}

\title{Coreless vorticity in multicomponent Bose and Fermi superfluids}

\author{G. Catelani}
\altaffiliation[Present address: ]{Department of Physics, Yale University, New Haven, Connecticut 06520, USA}
\author{E. A. Yuzbashyan}
\affiliation{Center for Materials Theory, Department of Physics and Astronomy,
Rutgers University, Piscataway, New Jersey 08854, USA}

\begin{abstract}

We consider quantized vortices in  two-component Bose-Einstein
condensates and three-component Fermi gases with attractive
interactions. In these systems, the vortex core can be either empty
(normal in the fermion case) or filled with another superfluid. We
determine critical values of the parameters -- chemical potentials,
scattering lengths and, for Fermi gases, temperature -- at which a
phase transition between the two types of vortices occurs.
Population imbalance can lead to superfluid core (coreless)
vorticity in multicomponent superfluids which otherwise support
only usual vortices. For multicomponent Fermi gases, we construct
the phase diagram including regions of coreless vorticity.  We
extend our results to trapped bosons and fermions using an
appropriate local approximation, which goes beyond the usual
Thomas-Fermi approximation for trapped bosons.

\end{abstract}
\date{\today}
\maketitle

\section{Introduction}

Properties of quantized vortex cores in superfluids have been actively  researched for many
years. For example, it was realized \cite{CDM} that in type II superconductors there are low-energy
states  bound to the core and their effect on the local density of states
was   studied experimentally~\cite{vscexp} and theoretically \cite{vscth}. More recently,
observations of quantized vortices provided key evidence for superfluidity in both
single component Bose-Einstein condensates (BECs) and two-component superfluid Fermi gases,
where vortex cores are detected as regions of suppressed particle
density \cite{varrbec,varrfg}.  The core states in Fermi gases were considered
in \cite{vortBECBCS}.

The situation in unconventional superfluids is more complex. For instance,
in superfluid $^3$He-B the vortex core is in a ferromagnetic superfluid state \cite{vHe3}.
In high-temperature superconductors with $d$-wave order parameter
an $s$-wave component must be present  in the core \cite{dswave}.
In a color superconductor --  high-density, low-temperature
quark matter -- there are many distinct fermion species (components) that can pair up, leading
to two types of vortices -- Abelian and non-Abelian \cite{colvor}.
Similarly, spinor atomic BECs having several bosonic components host, in addition to vortices,
other types of topological excitations such
as hedgehogs and skyrmions \cite{BECrev}. This cannot happen, however,
if the symmetry between the components is explicitly broken, in which case only vortices are possible.

\begin{figure}
\begin{flushleft}\includegraphics[width=0.48\textwidth]{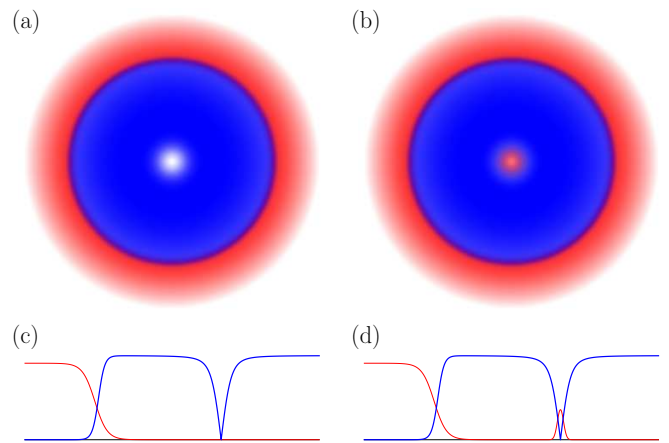}\end{flushleft}
\caption{Order parameters for normal-core [panels (a) and (c)] and superfluid-core
[(b) and (d)] vortices in multicomponent BECs and Fermi gases. Colors distinguish different superfluids. Top row:
 qualitative behavior of the order parameters in a plane of a trapped superfluid.
Color intensity is proportional to the local value of the order parameters. Bottom row: profiles of
the condensate wave functions in the absence of trapping.}
\label{fig:0}
\end{figure}

Consider, for example, a
three-component Fermi gas, which can support three distinct superfluid states \cite{ptm, CY}, with
a vortex in superfluid $S_1$. The core of the vortex can be in the normal state, as for a two-component
Fermi gas, or it can condense into one of the two other superfluid states -- see Fig.~\ref{fig:0}. In view of
current efforts to achieve superfluidity in three-component Fermi gases \cite{3fge1, 3fge2},
it is important to understand which of these scenarios is realized and/or how to drive the
transition between normal- and superfluid-core vortices.
Similarly, competition between empty- and filled-core vortices takes place in a two-component BEC (2BEC).
In fact, filled-core vortices have been experimentally observed \cite{vortexp, vortexp2}, and their properties
are in the focus of numerous theoretical studies \cite{stabcond1,stabcond2,tfavort,binary,tfaempty}
-- see \ocite{BECrev} for a review.

In this paper, we study vortices in 2BECs and  three-component Fermi
gases. We show that both empty-(normal-) and filled-(superfluid-)core vortices can be
realized depending on the parameters characterizing the bosonic (fermionic) system: chemical
potentials, scattering lengths, and temperature in the case of
fermions. We first focus on 2BEC at zero temperature and derive,
using Gross-Pitaevskii equations, the critical relationship
between the parameters of the system which determines whether the
vortex is empty or filled. Fermions at sufficiently high temperature
can be analyzed in a similar manner with the help of the
Ginzburg-Landau expansion. This enables us to find a condition for
the normal- to superfluid-core transition for vortices in a
three-component Fermi gas. In particular, for balanced systems with
equal populations of three fermion species, we explicitly obtain the
critical temperature for this transition. We first consider bosons
and fermions in the absence of an external potential, then extend
our results to trapped systems. We further discuss the validity of
the widely used Thomas-Fermi and local density approximations for
analyzing vortex cores in trapped condensates.

\section{Two-component Bose system}
\label{sec:bec}

The low temperature properties of a BEC are well described in terms of a
condensate wave function $\Psi$ which obeys the Gross-Pitaevskii equation, which captures the
effects of interactions at the mean field level. First, let us treat a 2BEC free of the trapping potential.
The coupled Gross-Pitaevskii  equations for the wave functions $\Psi_i$, $i=1,2$,
of the two condensates in a stationary state can be obtained by minimizing
the following energy density functional \cite{BECrev}:
\be\label{egp}\begin{split}
E = & \sum_{i=1}^2\left[ \frac{1}{2m_i} |\nabla\Psi_i|^2 - \mu_i|\Psi_i|^2 +
\frac{u_{ii}}{2}|\Psi_i|^4\right] \\
& + u_{12}|\Psi_1|^2|\Psi_2|^2,
\end{split}\ee
where $m_i$ and $\mu_i$ are the atomic masses and  chemical potentials, respectively. The coupling constants $u_{ij}$ are
related to the (positive) scattering lengths $a_{ij}$
\be\label{scat}
u_{ij} = 2\pi a_{ij}
\left(1/m_i+1/m_j\right).
\ee

We consider the case in which the two BECs phase separate (i.e., $u_{12}^2-u_{11}u_{22} > 0$)
and assume that there is a vortex line along the $z$ axis in condensate 1. Then the wave
function of the condensate has the form
\be\label{vorp1}
\Psi_1(\mathbf{r},\phi) = \sqrt{\frac{\mu_1}{u_{11}}} e^{i\phi} f\left(\frac{r}{\xi_1}\right),
\ee
where $\xi_1=1/\sqrt{2m_1\mu_1}$ is the healing length and $\mathbf{r}=(r,z)$.
The profile function $f(x)$ is the solution of the non-linear differential equation
(primes denote derivatives)
\be\label{vpe}
f'' +\frac{f'}{x}-\frac{f}{x^2}= - f+ f^3
\ee
with the boundary conditions $f(0)=0$ and $f(x)\to 1$ as $x\to\infty$.

\begin{figure}[!t]
\begin{flushleft}\includegraphics[width=0.47\textwidth]{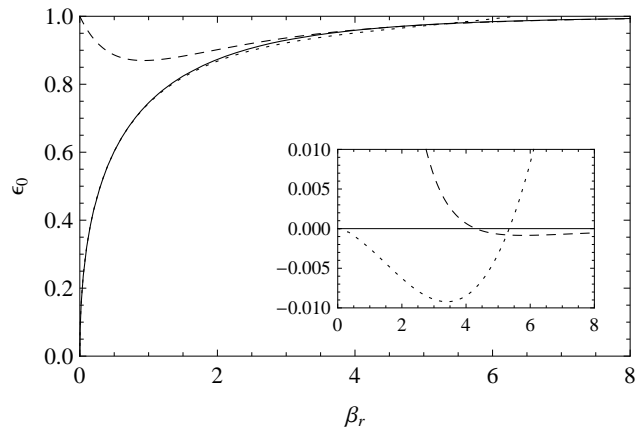}\end{flushleft}
\caption{Ground state energy $\epsilon_0$ of Schr\"odinger \eref{sche} as a function of the inverse mass $\beta_r$.
Solid line: numerical calculation. Dotted (dashed) line: analytical approximation
\eref{e0apps} for $\beta_r \lesssim 3$ ($\beta_r \gtrsim 3$). Inset: relative deviations between the approximations and the numerical results.}
\label{fig:e0br}
\end{figure}

Condensate 2 fills the core of the vortex in condensate 1 when this is energetically favorable.
At the transition from and empty ($\Psi_2=0$ in the core) to a filled core ($\Psi_2\neq 0$),
$\Psi_2$ is infinitesimally small. The energy difference between filled and
empty core states to the lowest order in $\Psi_2$ is
\be\label{de}
\delta E = L_z\!\int\!d^2r \left[ \frac{1}{2m_2} |\nabla \Psi_2|^2
+u_{12} |\Psi_1|^2|\Psi_2|^2  -\mu_2 |\Psi_2|^2 \right],
\ee
where $L_z$ is the size of the system in the $z$ direction.
We can neglect the effect of $\Psi_2$ on $\Psi_1$, since it is of higher order in $\Psi_2$.
This equation can be rewritten as
\be\label{de2}
\delta E = L_z\xi_1^2 \left[ \frac{u_{12}}{u_{11}}\epsilon_0(\beta_r)\mu_1 - \mu_2 \right]
\int\!d^2\rho \, |\Psi_2|^2 \, ,
\ee
where $\epsilon_0$ is the (dimensionless) ground
state energy of the following two-dimensional Schr\"odinger equation:
\be\label{sche}
-\beta_r \nabla^2 \Psi_2(\rho,\phi) + f^2(\rho) \Psi_2(\rho,\phi) = \epsilon_0 \Psi_2(\rho,\phi)
\ee
with $\rho=r/\xi_1$ and the ``inverse mass'' being
$
\beta_r= m_1 u_{11}/m_2 u_{12}
$.
The analysis reported in Appendix~\ref{appA}
shows that a good approximation for the
monotonically increasing function $\epsilon_0(\beta_r)$ is given by (see also Fig.~\ref{fig:e0br})
\be\label{e0apps}
\epsilon_0 (\beta_r) \simeq \left\{
\begin{array}{lcl}
2\sqrt{\beta_r}c_0 -\frac{1}{2}\beta_r + \frac{16c_0^2-4}{32c_0}\beta_r^{3/2}, & & \beta_r \lesssim 3, \\
1 -4 e^{-2\gamma_E}c_0^2 \beta_r e^{-2\sqrt{\beta_r}\arctan \left[c_0^2 2\sqrt{\beta_r}\right]},
& & \beta_r \gtrsim 3,
\end{array}
\right.
\ee
where $\gamma_E$ is Euler's constant and $c_0=f'(0)\simeq 0.58319$.

As an example, consider a system with $m_1=m_2$ and $\mu_1=\mu_2$. Then \esref{de2} and \rref{e0apps}
show that for $a_{22} < a_{12} < a_{11}$
a coreless vortex in superfluid 1 is more stable than a coreless vortex in superfluid 2, since the
latter state has higher energy.   This
is  in agreement with the experimental findings of \ocite{vortexp}.

More generally,
the sign of the term in square brackets in \eref{de2} determines the stable core state. If it is positive,
it costs energy to fill the core. Hence the core  is empty if
\be\label{emptycond}
\frac{u_{11}}{u_{12}} < \frac{\mu_1}{\mu_2}\epsilon_0 \left(\frac{m_1 u_{11}}{m_2 u_{12}}\right).
\ee
In other words, \eref{emptycond} defines a
surface in the space of parameters that separates the regions of empty and filled core vorticities.
For example, we see that population imbalance, which modifies the ratio $\mu_1/\mu_2$, impedes
coreless vorticity in one of the condensates while favoring it in the other, as seen by
exchanging $1\leftrightarrow 2$. Analogous conclusions hold for the ratio between intra- and
interspecies scattering length and the ratio of masses. Interestingly,
while our calculations are performed in the thermodynamic limit,
qualitatively similar conclusions were reached in Refs.~\onlinecite{stabcond1}-\onlinecite{stabcond2},
where stability conditions were derived for vortices in a small trapped condensate.

Now, let us analyze the effects due to an external potential. For simplicity,  we
consider a  spherically symmetric harmonic trap $V(\mathbf{r})$
and assume, without loss of generality,
that condensate 1 occupies its center, while superfluid 2 forms a shell around it. Vortices in trapped 2BEC have been previously studied in Refs.~\onlinecite{tfavort,binary,tfaempty} using the
Thomas-Fermi approximation (TFA).
In this approximation only the angular part of the kinetic energy terms
in \esref{egp} and \rref{de} is kept.  Then, the wave functions $\Psi_{i}$, $i=1,2$ of
the condensates are obtained from those in the absence of the trapping potential via the
replacement $\mu_i \to \mu_i(\mathbf{r}) \equiv \mu_i(0) - V(\mathbf{r})$, i.e., the trapped 2BEC is taken to be locally
uniform. This is justified in large condensates, $R_i \gg \xi_i$, where $R_i$ and $\xi_i$ are condensate
sizes and healing lengths, respectively.

Neglecting the radial parts of the kinetic energy,  however, is
a good approximation only when $\Psi_{i}$
varies over distances much larger than $\xi_i$. As we have seen above, this is not the case close to the
vortex line, where both $\Psi_1$ and $\Psi_2$ vary on the scale $\xi_1$, see Eqs.~(\ref{vpe}) and
\rref{sche}. We therefore expect the TFA to break down in determining the state in the vortex core. Indeed,
one of its artifacts is that $\Psi_1$  is identically zero in a finite region inside the
core \cite{tfavort,tfaempty}. This implies that the second term in \eref{de} vanishes too, while the
last term always makes it energetically favorable for the otherwise empty core to
be filled by the second superfluid. Thus, the question whether the core
is empty or filled cannot be resolved within the TFA, prompting arbitrary assumptions of
filled \cite{tfavort,binary} and empty \cite{tfaempty} core vortices in the literature.

This question,  as we now show, can be accurately answered for large condensates by
combining our approach with the local uniformity assumption discussed above. For example, let us work
out the condition under which an empty core vortex is realized in superfluid 1 for $m_1=m_2$.
The separation of scales, $R_i \gg \xi_i$, allows us to describe the vortex profile by \eref{vorp1}
with $\mu_i \to \mu_i(\mathbf{r})$. In other words, we neglect gradients
of $\mu_1(\mathbf{r})$ but not of $f(r)$. This approximation is valid at distances $L\gg \xi_1$ from the
interfaces between the condensates \cite{becint}. In a plane perpendicular to the vortex line, the wave
functions$\Psi_i$ looks like Figs.~\ref{fig:0}(c) and (d), except for the curvature imposed by
the trapping potential.  $\Psi_1(\mathbf{r},\phi)$  changes rapidly near the core as $r$ varies over
distances $r\sim \xi_1$ and is smooth on this scale when moving along
the $z$ axis (vortex line).

Therefore,  the stability of an empty core vortex for any $z$
is determined by \eref{emptycond}
with $\mu_i \to \mu_i(r_c,z)$,  $r_c$ being the position of the core. We obtain
\be\label{nec}
\frac{\mu_2(r_c,z)}{\mu_1(r_c,z)} <
\frac{a_{12}}{a_{11}}\, \epsilon_0\!\left(\frac{ a_{11}}{ a_{12}}\right),
\ee
where we took into account $m_1=m_2=m$ and $u_{ij}=4\pi a_{ij}/m$ -- see \eref{scat}. Further, one can
show that $\mu_2(0)>\mu_1(0)$ is necessary for superfluid 1 to occupy the center of the
trap \cite{tfaempty,note1}. It follows that  the left-hand side of \eref{nec} is a monotonically increasing
function of  $|z|$, i.e., it reaches its maximum at the interface
between the two condensates. The position $\mathbf{r}_{in}$ of the interface
is determined from the condition $\mu_2(\mathbf{r}_{in})/\mu_1(\mathbf{r}_{in})=\sqrt{a_{22}/a_{11}}$
obtained by equating the pressures on the two sides \cite{tfaempty,note1},
so that \eref{nec} becomes
\be\label{tfacons}
\sqrt{\frac{a_{22}}{a_{11}}}
< \frac{a_{12}}{a_{11}}\,\epsilon_0\!\left(\frac{a_{11}}{a_{12}}\right).
\ee
If this inequality holds,  it costs energy to introduce the second condensate everywhere along
the vortex core and the empty vortex is
the stable state. Otherwise, the core is partially or fully filled.  Since the left-hand side
of \eref{nec} is larger at the interface, when varying the scattering lengths
the core will be filled starting from the interface between the condensates towards the trap center.
Using \eref{e0apps} and the experimental values reported in \ocite{vortexp}, we find that
condition \rref{tfacons} is violated in this experiment. This invalidates the approach
of \ocite{tfaempty} for the description of this experiment based on an empty core assumption.

A few comments are in order about the present derivation.
We have not used anywhere the condition $a_{12}^2-a_{11}a_{22} > 0$ that ensures phase separation in
the absence of external potentials.
For trapped 2BEC, phase separation can occur as long as $a_{12} >0$,
although if $a_{12}^2-a_{11}a_{22} < 0$ a coexistence region is present \cite{binary}.
In this case our analysis applies at distances $L\gg \xi_i$ from the coexistence region,
but the inequality \rref{tfacons} is always violated since $\epsilon_0 < 1$ -- see \eref{e0apps}.
This means that empty vortices are possible only in phase-separated condensates.
More generally,
as discussed above, to obtain \eref{tfacons} we use locally the result derived in the
thermodynamic limit. This approximation is valid assuming slow variation of the wave functions
along the vortex line.
This assumption breaks down near the interface between the two phases, and
even in fully phase-separated 2BEC the regions where
the vortex core meets the interface are beyond the reach of the present approach.

\section{Three-component Fermi system}
\label{sec:fg}

The treatment presented above can be applied to the study of a vortex
in a multicomponent Fermi system at sufficiently high temperature. Consider, in
particular,   a three-component system at weak coupling
near second order phase transition lines. The Ginzburg-Landau (GL) expansion for the thermodynamic potential
$\Omega$ is \cite{CY}
\begin{eqnarray}
\frac{\Omega-\Omega_N}{\nu} & =
& \sum_{i=1}^2\left\{ \alpha_i |\Delta_i|^2 +\frac{\beta_{ii}}{2}\left[|\Delta_i|^4 +
\frac{v_F^2}{3}|\nabla \Delta_i|^2\right]\right\} \nonumber \\
&& +\beta_{12} |\Delta_1|^2|\Delta_2|^2 ,
\label{GLe}
\end{eqnarray}
where $\Omega_N$ is the normal-state potential, $\nu$ is the density of states at the Fermi energy,
$v_F$ is the Fermi velocity, and $\Delta_i$ is the order parameter describing pairing of
particles belonging to species $j$ and $k$ with $j$, $k$, and $i$ all different.

The coefficients $\alpha_i$, $\beta_{ij}$ in \eref{GLe} are
\begin{eqnarray}
h_1 &=& \mu_3 - \mu_2 \, , \quad h_2 = \mu_3 - \mu_1 \, , \nonumber \\
\alpha_i &=& \ln\frac{T}{T_{c_i}} + \Re\,\psi\left(\frac{1}{2} +  \frac{ \mathrm{i}h_i}{4\pi T}\right)
- \psi\left(\frac{1}{2}\right) ,
\\
\beta_{ij}  &=& \frac{1}{h_i-h_j} \frac{1}{4\pi T}
\Im \bigg[\psi'\left(\frac{1}{2}+\frac{\mathrm{i}h_j}{4\pi T}\right)
\bigg]
+ (j \to i),
\nonumber
\end{eqnarray}
where $\mu_i$ are  the chemical potentials of the three species, $\psi$ is the digamma function,
$\beta_{ii}$ is obtained from $\beta_{ij}$ in the limit $h_j \to h_i$, and $T_{c_1}$ ($T_{c_2}$) is the critical temperature for the normal-superfluid transition for a two-component gas of species 2 and 3 (1 and 3). With no loss of generality we assume $T_{c_1}>T_{c_2}$.
For simplicity, we neglect the weakest of the three possible interactions, say between species 1 and 2 (recall that due to Pauli exclusion principle only interspecies scattering is possible in the $s$-wave channel). Then, there are
only three main phases of the three-component system -- the normal phase N ($\Delta_1=\Delta_2=0$), superfluid
$S_1$ ($\Delta_1\ne0, \Delta_2=0$), and superfluid
$S_2$ ($\Delta_1=0, \Delta_2\ne0$).
The applicability of the GL expansion requires
$T\gtrsim 0.56 T_{c_1}$ and $|h_i|/4\pi T\lesssim 0.30$.
In this regime $\beta_{ij}>0$, indicating repulsion (this is a consequence of Pauli exclusion),
and $\beta_{12}^2-\beta_1\beta_2\ge0$, leading to phase separation between the superfluid states.
The first-order $S_1$-$S_2$ transition is accessible within the GL description
when $T_{c_1}-T_{c_2}\ll T_{c_2}$ and $T_{c_2}-T \ll T_{c_2}$ -- see \ocite{CY} for more details.

Comparison of \esref{egp} and \rref{GLe} shows that, up to a redefinition
of the various parameters, the energy argument
discussed in  Sec.~\ref{sec:bec}
can be applied to the thermodynamics of the three-component Fermi gas.
For example, a vortex in superfluid $S_1$ is described by the order parameter [cf. \eref{vorp1}]
\be
\Delta_1(\mathbf{r},\phi) = \sqrt{\frac{-\alpha_1}{\beta_{11}}} e^{i\phi} f\left(\frac{r}{\xi_1}\right)
\ee
with the coherence length $\xi_1 = v_F \sqrt{\beta_{11}/6(-\alpha_1)}$.

Repeating the previous analysis for such a vortex we find the following
condition for the second-order phase transition
between normal and superfluid core (or equivalently between standard and coreless vortex):
\be\label{ecf}
\alpha_2 - \alpha_{1} \frac{\beta_{12}}{\beta_{11}}
\,\epsilon_0\!\left(\frac{\beta_{22}}{\beta_{12}}\right) = 0,
\ee
where the function $\epsilon_0(x)$ is given by \eref{e0apps}.
The first term in \eref{ecf}  originates from the
energy gained by condensation of the originally uncondensed species.
The condensation takes place in the core region, where pairs of the first superfluid are broken due to
vorticity.
The second term is due to the energy costs associated with deforming
the order parameter $\Delta_2$ [kinetic energy in \eref{sche}] and the repulsive interaction between the
two superfluids (potential energy).

\begin{figure}
\begin{flushleft}\includegraphics[width=0.45\textwidth]{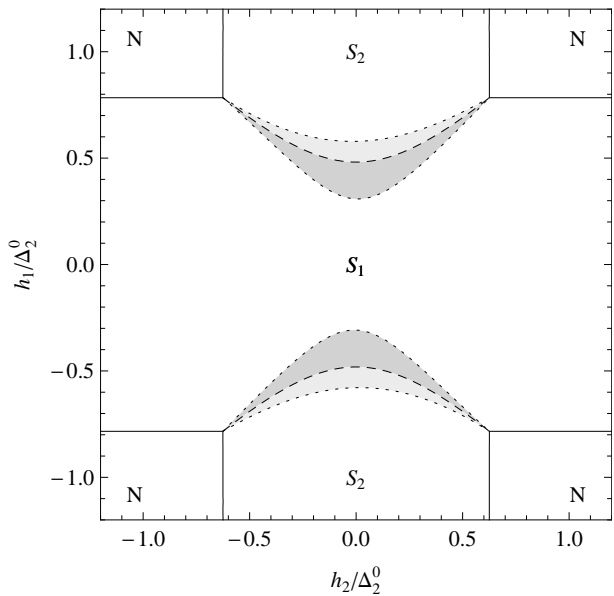}\end{flushleft}
\caption{High temperature ($T=0.93 T_{c2}$) phase diagram for a three-component Fermi gas
with $T_{c_1}/T_{c_2}=1.04$ in the plane of chemical potential differences $h_i$.
Solid lines: second-order normal-superfluid phase transitions. Dashed lines: first-order
transition between the two superfluid states. In dark grey are the regions
where the core of a vortex in the $S_1$ superfluid is filled by the $S_2$ superfluid.
Light grey: $S_1$ superfluid fills the $S_2$ vortex core.}
\label{fig:2}
\end{figure}

\begin{figure}
\begin{flushleft}\includegraphics[width=0.45\textwidth]{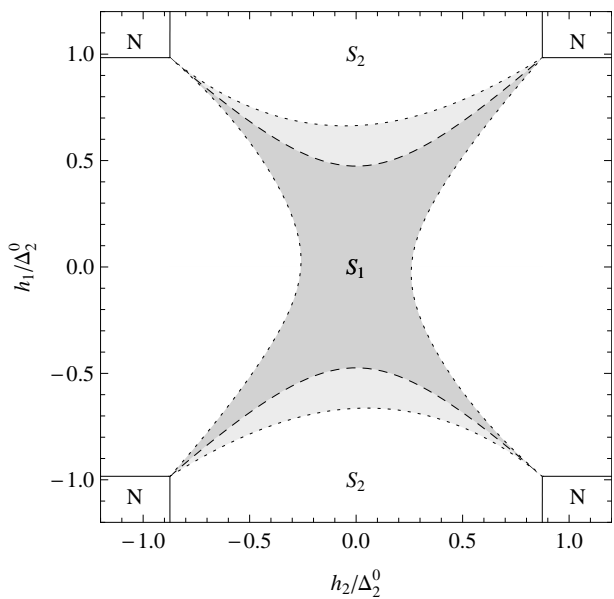}\end{flushleft}
\caption{Phase diagram as in Fig.~\ref{fig:2} but for a lower temperature ($T=0.85 T_{c2}$). Note that
now the coreless vortex state is present even for balanced gases with $h_i=0$.}
\label{fig:3}
\end{figure}

At a given temperature $T<T_{c_2}$, \eref{ecf} determines the area in the
$h_1$-$h_2$ space where the core is filled. It can be satisfied only in the central region of this
space where both coefficients $\alpha_{1,2}$ are negative, i.e., when condensation is in principle possible in
both channels. We show two examples of phase diagrams in Figs.~\ref{fig:2}
and \ref{fig:3}. At a temperature close to $T_{c_2}$ (Fig.~\ref{fig:2}) the core is filled only in
small regions around the first order phase transition between the two superfluid states.
As the temperature is lowered (Fig.~\ref{fig:3})
the regions where superfluidity is possible expand, and so do the regions of coreless vorticity.
In particular, as the temperature decreases, coreless vortices become possible even in balanced
systems with equal populations ($h_1=h_2=0$). In this case the coefficients
$\beta_{ij}$ all coincide and from \eref{ecf} we find that the temperature $T_o$ for the onset
of coreless vorticity is related to the superfluid critical temperatures as
\be\label{to}
T_o = T_{c_2} \left(\frac{T_{c_2}}{T_{c_1}}\right)^\gamma, \quad
\gamma =\frac{\epsilon_0(1)}{1-\epsilon_0(1)} \simeq 2.92 \, .
\ee
Above $T_o$ the core is in a normal state, while for $T<T_o$ it is superfluid.

\eref{ecf} can be used to study trapped Fermi
gases within the local density approximation (LDA) so long as condensate size $R$
is large compared to the coherence length $\xi_1$. Then, we can simply substitute local
values of the critical temperatures $T_{c_i}(r)$ into \eref{ecf} (note that chemical potential differences $h_i$ are position independent within LDA).
The condition $R\gg \xi_1$ is satisfied if the total particle number $N_t$ is large enough.
 For a somewhat weak
interaction $k_F |a_s| \simeq 0.5$ the condition on the particle number is $N_t \gg 10^3$, see \ocite{CY}. We remind,
however, that
for imbalanced systems the presence of domain walls could require larger particle numbers for
the LDA to be valid.

The above calculations are valid in the weak
coupling regime near second order phase transitions.
The latter requirement is satisfied for all relevant chemical potential differences
if $T_{c_1}-T_{c_2}\ll T_{c_2}$ and $T$ is close to $T_{c_2}$.
Relaxing these conditions will not alter the qualitative picture, although it will affect
the quantitative results. For example, if the difference between the critical temperatures, $T_{c_1}-T_{c_2}$,
is large, we expect the actual onset temperature $T_o$ to be smaller than that predicted by
\eref{to}: at low temperature $T\to 0$ the order parameter in the core region rises on a length
scale of the order of the Fermi wave length \cite{vortBECBCS,ltv} rather than the much longer
(at weak coupling) coherence length.
At lowest order this is equivalent to an increase of $c_0$ in, e.g., \eref{e0apps}, which
would lead to a higher value for $\epsilon_0(1)$ in \eref{to} and hence a lower onset temperature.

\section{Summary}

We considered the empty- and filled-core vortex states in
two-component BECs described by the Gross-Pitaevskii equations, as well as the normal- and
superfluid-core vortices in three-component Fermi gases using a Ginzburg-Landau approach.
In the absence of external potentials, we derived the conditions
[\eref{emptycond} and \rref{ecf} for Bose and Fermi systems, respectively] that determine
the transition between the two states in terms of chemical potentials, scattering lengths and,
for Fermi gases, temperature. In particular, we obtained a simple expression, \eref{to}, for the onset
temperature of coreless vorticity in the population-balanced Fermi gas.
We also established when the vortex core remains empty in trapped Bose systems, see \eref{tfacons},
using a local approximation which goes beyond the usual Thomas-Fermi Approximation. We showed that, for equal masses of the
components, empty vortices are possible only in phase-separated 2BECs and that in partially filled
vortices the superfluid part of the core is in the region closer to the interface between the two
condensates. We similarly applied our findings to trapped multicomponent Fermi gases within the Local Density Approximation
and discussed  the limits of validity of this approach. The detailed study of the
superfluid core is left to future work, as is the extension of our results to Fermi-Bose mixtures.

\begin{acknowledgments}

This research was financially supported in part by the National Science
Foundation under Award No. NSF-DMR-0547769 and
the David and Lucille Packard Foundation.

\end{acknowledgments}

\appendix
\section{Calculation of the ground-state energy}
\label{appA}

In this Appendix
we present the calculation of the ground-state energy $\epsilon_0$ of \eref{sche}.
The potential term $f^2(\rho)$ is not known explicitly, except for its behavior at small and large
$x$. Indeed, near the origin
a power series for $f(\rho)$ can be found in term of one unknown parameter, $c_0$, which is
calculated using the boundary condition at infinity. The first few terms in this expansion are:
\be\label{f0lim}
f(\rho) = c_0 \rho - \frac{c_0}{8}\rho^3+\frac{c_0(1+ 8 c_0^2)}{192}\rho^5+O(\rho^7) \, .
\ee
The coefficient $c_0\simeq 0.58319$ can be evaluated with great precision
either numerically or with analytical methods \cite{c0ref}. As $\rho\to +\infty$ the
asymptotic expansion is:
\be\label{fasym}
f(\rho) = 1-\frac{1}{2\rho^2} - \frac{9}{8\rho^4} + O(\rho^{-6}) \, .
\ee
By appropriate rescalings, we show that this is
sufficient to obtain analytical estimates for $\epsilon_0$ at small and large $\beta_r$ which, moreover,
provide accurate estimates even at intermediate values.

\subsection{Small $\beta_r$}

The limit $\beta_r \to 0$ corresponds to a particle with large mass; therefore, its ground state
wave functions does not extend far from the origin.
This enables us to calculate $\epsilon_0$ perturbatively, using the harmonic oscillator as
starting point. Indeed,
after the rescaling $\rho = \beta_r^{1/4} c_0^{-1/2} x$, \eref{sche} becomes
\be\label{sche2}
-\nabla^2 \Psi_2 + \left[x^2 + V_{\beta_r}(x)\right] \Psi_2 = \tilde{\epsilon}_0 \Psi_2
\ee
with $\tilde{\epsilon}_0 = \epsilon_0/\sqrt{\beta_r}c_0$ and $V_{\beta_r}(x)$ the perturbation potential.
The latter is defined as the potential term of \eref{sche}, $f^2(\rho)$,
minus the harmonic part. From \eref{f0lim} we find the first few terms in the small-$x$ expansion:
\be
V_{\beta_r} (x) = -\sqrt{\beta_r}\frac{1}{4c_0}x^4+\beta_r\frac{5+16c_0^2}{192c_0^2}x^6 +
O(\beta_r^{3/2}x^8) \, .
\ee

Consistently with the truncation of the potential $V_{\beta_r}$, we calculate
$\tilde{\epsilon}_0$ to second order in the small parameter $\sqrt{\beta_r}$ via
standard time-independent perturbation theory \cite{LLqm}.
Since the potential does not break 2D rotational symmetry, to calculate the correction to the
ground state energy we only need to know the $s$-wave eigenvalues $\varepsilon_n$ and
eigenfunctions $\psi_n$ of the 2D harmonic oscillator:
\be\begin{split}
&\varepsilon_n  = 2 (2n+1) \, , \\
& \psi_n(x) = \sqrt{2} e^{-x^2/2} L_n(x^2) \, ,
\end{split}\ee
where $L_n$ are the Laguerre polynomials. In calculating the matrix elements of the perturbation
we use the identity
\be
u^n = n! \sum_{i=0}^{n} (-1)^i \frac{n!}{(n-i)!i!} L_i(u) \,
\ee
which follows by induction from the recurrence relation for the Laguerre polynomials \cite{absteg}.
After straightforward algebra we find
\be\label{e0pt}
\epsilon_0 = 2\sqrt{\beta_r}c_0 -\frac{1}{2}\beta_r + \frac{16c_0^2-4}{32c_0}\beta_r^{3/2} + O(\beta_r^2) \, .
\ee
Requiring the second term to be a small correction gives the condition $\beta_r\ll 5.44$; hence
the expansion should be reliable up to $\beta_r$ of order 1. Indeed, both the
calculation of the coefficient of the $\beta_r^2$ term ($\simeq 0.003)$ and comparison
with numerics (see the end of this
Appendix)
show that \eref{e0pt} is still a good approximation even for $\beta_r \gtrsim 1$.

\subsection{Large $\beta_r$}

In the limit of large $\beta_r$ the kinetic term in \eref{sche} becomes dominant, which
physically correspond to an almost free particle, and the (properly defined) potential $U(\rho)$
can be treated as a shallow one. Since $f^2(\rho) \to 1$ at large $\rho$, we define
$U(\rho) = f^2(\rho)-1$ and
$\varepsilon = 1-\epsilon_0$, so that positive energies correspond to bound states.
In two dimensions the ground state energy in a shallow potential depends exponentially
on the inverse mass \cite{LLqm}:
$\varepsilon \propto \beta_r e^{-c \beta_r}$, where the constant $c\propto |\int d^2\rho \, U(\rho)|^{-1}$.
This result holds when the integral converges, while in our case
the integral is logarithmically divergent, see \eref{fasym}.
We can adapt to the present situation the derivation of the above estimate for $\varepsilon$
presented in \ocite{LLqm}, \S 45.
We show at the end of this appendix that the expression thus obtained agrees well with numerical
calculations.

We define
\be
y = \sqrt{\frac{\varepsilon}{\beta_r}}\rho
\ee
and rewrite \eref{sche} as
\be\label{sche3}
\frac{1}{y}\left[y\Psi_2'(y)\right]' = U\left(y\right)\Psi_2(y)
\ee
with
\be
U(y) = 1+\frac{1}{\varepsilon}\left[f^2\left(\sqrt{\frac{\beta_r}{\varepsilon}}y\right)-1\right] .
\ee
Here we include the eigenvalue term ($\propto\varepsilon$ in the original equation)
into the potential energy $U$. In a small circle around the origin
of radius $y_0$ the potential can be approximated as
\be\label{Uorex}
U(y)\simeq -\frac{1}{\varepsilon} \, .
\ee
In this region the wave function can be taken as approximately constant, $\Psi_2 = 1$, and integrating
both side of \eref{sche3} gives
\be\label{ldleft}
y_0 \Psi_2'(y_0) = -\frac{1}{2\varepsilon}y_0^2 \, .
\ee
To estimate $y_0$, we note that
according to \eref{f0lim} the length scale over which the potential varies
appreciably near
the origin is $x\sim 1/c_0$, so we take $y_0 = \sqrt{\varepsilon/\beta_r}/c_0$.

For $y\gg \sqrt{\varepsilon/\beta_r}$, the potential $U$ is approximately
\be
U(y)\simeq 1-\frac{1}{\beta_r y^2}
\ee
and \eref{sche3} has as solution the modified Bessel function of imaginary order
$K_{i/\sqrt{\beta_r}}(y)$. We want to match its logarithmic derivative to the estimate in \eref{ldleft};
to do so, we use the following approximate expression valid for $y, 1/\sqrt{\beta_r} \ll 1$ \cite{bessim}
\be\label{Kiapp}
K_{i/\sqrt{\beta_r}}(y) \simeq -\sqrt{\beta_r} \sin\left[\frac{1}{\sqrt{\beta_r}}\left(
\log \frac{y}{2} + \gamma_E \right)\right].
\ee
Direct inspection of the matching condition at $y_0$ suggests taking $\varepsilon$ in the form
\be\label{veg}
\varepsilon = 4 e^{-2\gamma_E}c_0^2 \beta_r e^{-b\sqrt{\beta_r}}
\ee
for some parameter $b$. Then the matching condition reduces to
\be
\frac{1}{\sqrt{\beta_r}\tan \left[b/2\right]} = \frac{1}{2c_0^2\beta_r}
\ee
and solving for $b$ we find
\be
b = 2\arctan \left[2c_0^2\sqrt{\beta_r}\right] \, .
\ee
Substituting this back into \eref{veg} we finally arrive at
\be\label{e0lb}
\epsilon_0 = 1 -4 e^{-2\gamma_E}c_0^2 \beta_r e^{-2\sqrt{\beta_r}\arctan \left[c_0^2 2\sqrt{\beta_r}\right]}
\, .
\ee

Note that, in agreement with \eref{Kiapp}, $K_{i/\sqrt{\beta_r}}(y)$ has an infinite number of zeros
at (approximate) positions
$y_n = 2e^{-\pi n\sqrt{\beta_r} - \gamma_E}$, $n=1,2,3, \ldots$, and no other zeros at $y>y_1$.
It is easy to check that $y_n < y_0$, so that the approximate wave function constructed in the course
of this derivation has no zeros, as expected for the ground state.

To check the accuracy of the estimates in \esref{e0pt} and \rref{e0lb}, we also solve \eref{sche} numerically.
That is, we first find a numerical solution to \eref{vpe} for $f(x)$; then we find numerical estimates
of the ground state energy of \eref{sche} for various $\beta_r$. Interpolation of these numerical results
gives the solid curve in Fig.~\ref{fig:e0br}, where we also plot \esref{e0pt} and \rref{e0lb} for comparison.
Using \eref{e0pt} for $\beta_r \lesssim 3$ and \eref{e0lb}
for $\beta_r \gtrsim 3$ gives estimates that deviate less than 1\% from the numerics -- see
the inset of Fig.~\ref{fig:e0br}.


\begin{thebibliography}{99}
\bibitem{CDM}C. Caroli, P. G. de Gennes, J. Matricon, Phys. Lett. {\bf 9}, 307 (1964).
\bibitem{vscexp}H. F. Hess {\it et al}., Phys. Rev. Lett. {\bf 62}, 214 (1989).
\bibitem{vscth}F. Gygi and M. Schl\"uter, Phys. Rev. Lett. {\bf 65}, 1820 (1990);
Phys. Rev. B {\bf 43}, 7609 (1991).
\bibitem{varrbec}M. R. Andrews \textit{et al}., Science {\bf 273}, 84 (1996).
\bibitem{varrfg}M. W. Zwierlein \textit{et al}., Nature {\bf 453}, 1047 (2005).
\bibitem{vortBECBCS}M. Machida and T. Koyama, Phys. Rev. Lett. {\bf 94}, 140401 (2005);
R. Sensarma, M. Randeria, and T.-L. Ho, Phys. Rev. Lett. {\bf 96}, 090403 (2006).
\bibitem{vHe3}M. M. Salomaa and G. E. Volovik, Rev. Mod. Phys. {\bf 59}, 533 (1987).
\bibitem{dswave}Y. Ren, J. Xu, and C. S. Ting, Phys. Rev. Lett. {\bf 74}, 3680 (1995).
\bibitem{colvor}A.P. Balachandran, S. Digal, and T. Matsuura,  Phys. Rev. D {\bf 73}, 074009 (2006).
\bibitem{BECrev}K. Kasamatsu, M. Tsubota, and M. Ueda, Int. J. Mod. Phys. B. {\bf 19}, 1835 (2005).
\bibitem{ptm}T. Paananen, P. Torma, and J.-P. Martikainen, Phys. Rev. A {\bf 75}, 023622 (2007).
\bibitem{CY}G. Catelani and E. A. Yuzbashyan, Phys. Rev. A {\bf 78}, 033615 (2008).
\bibitem{3fge1}T. B. Ottenstein \textit{et al}., Phys. Rev. Lett. {\bf 101}, 203202 (2008).
\bibitem{3fge2}J. H. Huckans \textit{et al}., Phys. Rev. Lett. {\bf 102}, 165302 (2009).
\bibitem{vortexp}M. R. Matthews \textit{et al}., Phys. Rev. Lett. {\bf 83}, 2498 (1999).
\bibitem{vortexp2}B. P. Anderson \textit{et al}., Phys. Rev. Lett. {\bf 85}, 2857 (2000).
\bibitem{stabcond1}D. V. Skryabin, Phys. Rev. A {\bf 63}, 013602 (2000).
\bibitem{stabcond2}J. J. Garcia-Ripoll and V. M. Perez-Garcia, Phys. Rev. Lett. {\bf 84}, 4264 (2000);
V. M. Perez-Garcia and J. J. Garcia-Ripoll, Phys. Rev. A {\bf 62}, 033601 (2000).
\bibitem{tfavort}D. M. Jezek, P.  Capuzzi, and H. M. Cataldo, Phys. Rev. A {\bf 64}, 023605 (2001).
\bibitem{binary}T.-L. Ho and V. B. Shenoy, Phys. Rev. Lett. {\bf 77}, 3276 (1996).
\bibitem{tfaempty}S. T. Chui, V. N. Ryzhov, and E. E. Tareyeva, Phys. Rev. A {\bf 63}, 023605 (2001);
JETP {\bf 91}, 1183 (2000).
\bibitem{becint}R. A. Barankov, Phys. Rev. A {\bf 66}, 013612 (2002); B. Van
Schaeybroeck, \textit{ibid}. {\bf 78}, 023624 (2008).
\bibitem{note1} Note that the presence of a vortex does not significantly alter the chemical potentials and
the intereface as it occupies  a small fraction $\sim(\xi_1/R_1)^2\ll1$ of the condensate.
Therefore, one can use the TFA to determine these quantities.
\bibitem{ltv}L. Kramer and W. Pesch, Z. Physik {\bf 269}, 59 (1974).
\bibitem{c0ref}N. G. Berloff, J. Phys. A: Math. Gen. {\bf 37}, 1617 (2004);
B. Boisseau \textit{et al}., J. Phys. A: Math. Theor. {\bf 40}, F215 (2007).
\bibitem{LLqm}L. D. Landau and E. M. Lifshitz, \textit{Quantum mechanics} (Butterworth-Heinemann, Oxford, 1981).
\bibitem{absteg}See, e.g., M. Abramowitz and I. A. Stegun (eds.), \textit{Handbook of Mathematical Functions}
(Dover, New York, 1964).
\bibitem{bessim}T. M. Dunster, SIAM J. Math. Anal. {\bf 21}, 995 (1990).
\end{thebibliography}
\end{document}